\documentstyle[12pt]{article}
\begin{document}
\baselineskip 24pt
\begin{center}
{\large\bf EVIDENCE FOR NEUTRINO BEING LIKELY A SUPERLUMINAL PARTICLE}
\end{center}

\vskip 24pt
\normalsize
\baselineskip 24pt
\begin{center}
GUANG-JIONG NI$^*$\footnote{$^*$ E-mail: gj\_ni@yahoo.com}\\
{\it Department of Physics, Fudan University, Shanghai 200433, China}\\
{\it Department of Physics, Portland State University, Portland, OR97207, USA
(Mailing address)}
\end{center}

\vskip 24pt
\noindent
{\bf Abstract} Experimental evidence and theoretical argument in
\baselineskip 24pt
favor the claim of neutrino being likely a superluminal
particle, a tachyon, are discussed from six aspects with details.

\newpage
\baselineskip 24pt
\noindent
{\bf 1. Introduction}

\vskip 24pt
\baselineskip 24pt
\indent The question of superluminal particle (tachyon) has a long history.
For a broad review on superluminal motion, see Racami's paper[1]. In this
paper we will concentrate on some crucial points which seem to obstruct the
general acceptance of the tachyon concept in physics community. They are:\\
\indent (a)Does the principle of the limiting speed deny the existence of a
tachyon?\\
\indent (b)Can the puzzle about the violation of causality in superluminal
motion be solved?\\
\indent (c)Can we find a quantum equation for tachyon and what is its
symmetry?\\
\indent (d)Could a possible tachyon be the neutrino? What is the crucial
evidence on this claim?\\
\indent (e)There is an analytical continuation of mass in the transition from
subluminal motion to superluminal one. What is its physical meaning?\\
\indent (f)Does the so-called Pseudotachyon[2] exist?\\
\indent We will discuss these questions in next six sections before the
summary and further discussion will be made in the final.

\vskip 48pt
\baselineskip 24pt
\noindent
{\bf 2. The Principle of Limiting Speed and the Principle of Causality}

\vskip 24pt
\baselineskip 24pt
\indent In the theory of special relativity (SR), the kinematic relation of a
particle reads:
\begin{equation}
E^2=\vec{p}^2c^2+m_0^2c^4
\end{equation}
where $E$, $\vec{p}$ and $m_0$ are energy, momentum and rest-mass of the
particle respectively. Then the velocity $\vec{u}$ of the particle can easily
be derived as:
\begin{equation}
\vec{u}=\vec{p}c^2/E,
\end{equation}
\begin{equation}
\vec{p}=\frac{m_0\vec{u}}{\sqrt{1-u^2/c^2}}, \hskip 0.3in
E=\frac{m_0c^2}{\sqrt{1-u^2/c^2}}.
\end{equation}
Obviously, the speed $u$ of particle cannot exceed that of light: $u<c$. In
other words, there is an upper limit $c$ for all (subluminal) particles with
nonzero rest-mass $m_0$.\\
\indent Because of the great success of SR, the existence of a limiting speed
gradually became a principle to construct new theory in physics. Sometimes, it
took an equivalent form as a principle of causality --- any particle, or
energy, or information can only be traveling in a speed smaller than (or at
most equal to) $c$ and along the time direction from the past to the future.\\
\indent By a careful reading on the original paper of Einstein in 1905, we
understand that there are two relativistic postulates for the establishment of
SR: the principle of relativity (A) and the principle of constancy of the
speed of light (B). The B is necessary to establish the Lorentz transformation
(LT) before the A can make sense in two (or more) inertial systems
($S$, $S'$ etc.) quantitatively. In other words, the light-speed $c$ exhibits
itself as a constant and serves as a means to fix the relationship of
coordinates between two systems. There is no need for $c$ being an upper limit
of particle velocity. The latter concept is merely a lemma of Eq.(3) which is
an outcome of SR rather than a starting-point of SR.\\
\indent Of course, the light-speed $c$ displays itself as a singular point in
the space-time as shown by the LT:
\begin{equation}
x'=\frac{x-vt}{\sqrt{1-v^2/c^2}}, \hskip 0.3in
t'=\frac{t-vx/c^2}{\sqrt{1-v^2/c^2}},
\end{equation}
where $v$ is the relative velocity between two systems $S$ and $S'$. Note that
$v$ always has an upper limit $c$ because we observers are composed of
ordinary particles obeying Eq.(3). However, on the other hand, being the
coordinates of some event, $x$ and $t$ can take arbitrary values on the
(two-dimensional) $x$-$t$ diagram.\\
\indent Let's consider a tachyon (P) with uniform velocity $u=x/t$ in the $S$
system. The particle's world-line OP is a straight line delivered from the
origin O and located below the line $x=ct$ (see [3-5]). Now consider another
$S'$ system moving with a velocity $v$ relative to $S$. Then the velocity of
tachyon in $S'$ will be $u'=x'/t'$. A strange phenomenon occurs as follows. In
accompanying with the increase of $v$, the $x'$ axis rotates counterclockwise
and acrosses the OP line at $v=c^2/u$. Just at this moment, the time
coordinate $t'$ of tachyon would become negative suddenly:
\begin{equation}
t'_p<0, \hskip 0.3in (u>c, \hskip 0.1in v>c^2/u).
\end{equation}
This was regarded as "tachton traveling backward in time" or a "violation to
the principle of causality" because the tachyon seems (in $S'$ system) to send
the information into the past! Many physicists pondered the above puzzle being
nonsense and believed even firmly the principle of limiting speed. So little
attention has been paid to discussions on the superluminal motion except in
some literatures, e.g., [1-5].\\
\indent Evidently, there is no way out in the pure classical theory. A
possible way out of this puzzle is to refer to the Stuekelberg-Feynman rule in
quantum electrodynamics --- a (negative-energy) electron traveling backward in
time is equivalent to a (positive-energy) positron traveling forward in time.
Then a "reinterpretation principle" was proposed by Sudarshan and Recami that
a (negative-energy) tachyon traveling backward in time may be viewed as an
(positive-energy) antitachyon traveling forward in time[1,6].\\
\indent Despite the attractiveness of above argument, we still need a rigorous
proof which can only be provided by a quantum equation for the tachyon. Before
we turn to the quantum realm, let's explore the above puzzle into a full
paradox[5].\\
\indent From the LT (4), the addition law for velocity can easily be derived
as:
\begin{equation}
u'=\frac{u-v}{1-uv/c^2}.
\end{equation}
There is a singular point for $v$ located at $c^2/u$ as long as $u>c$. Once
$v>c^2/u$, instead of Eq.(5), now we see that $u'$ changes its sign from
positive to negative:
\begin{equation}
u'<(-c), \hskip 0.3in (u>c^2/v, \hskip 0.1in or \hskip 0.1in v>c^2/u).
\end{equation}
However, Eq.(7) still remains as a puzzle because we have the momentum $p'>0$
as easily proved by the LT:
\begin{equation}
p'=\frac{p-vE/c^2}{\sqrt{1-v^2/c^2}}, \hskip 0.3in
E'=\frac{E-vp}{\sqrt{1-v^2/c^2}}.
\end{equation}
Now we write down the counterparts of Eqs.(1) and (3) for tachyon being:
\begin{equation}
E^2=p^2c^2-m_s^2c^4,
\end{equation}
\begin{equation}
p=\frac{m_su}{\sqrt{u^2/c^2-1}}, \hskip 0.3in
E=\frac{m_sc^2}{\sqrt{u^2/c^2-1}},
\end{equation}
where the real $m_s$ is named the proper mass or tachyon mass.\\
\indent We will derive (9) and (10) below. Now just combining (8) with (10),
we find:
\begin{equation}
p'=\frac{m_s(u-v)}{\sqrt{1-v^2/c^2}\sqrt{u^2/c^2-1}}>m_sc>0,
\end{equation}
\begin{equation}
E'=\frac{m_s(c^2-uv)}{\sqrt{1-v^2/c^2}\sqrt{u^2/c^2-1}}<0, \hskip 0.3in
(u>c^2/v, \hskip 0.1in or \hskip 0.1in v>c^2/u).
\end{equation}
Then new puzzle emerges: How can a particle have $u'<0$ whereas its $p'>0$?
What is the meaning of $E'<0$? (The energy must be positive definite in a
classical theory). We call all above puzzles, Eqs.(5), (7), (11) and (12) the
"superluminal paradox"[5].\\
\indent Evidently, to solve the superluminal paradox, we need more than a
"reinterpretation principle" as a generalization to the Stuekelberg-Feynman
rule. Even the latter needs a deeper explanation.

\vskip 48pt
\baselineskip 24pt
\noindent
{\bf 3. Dirac Equation Versus the Tachyon Equation}

\vskip 24pt
\baselineskip 24pt
\indent Obviously, we have to go beyond the realm of classical physics and
innovate new quantum theory for the tachyon. Then immediately, we are facing a
challenge: Do we already have a quantum theory for SR? If we don't have a
reliable quantum theory for SR, i.e., a new interpretation and derivation of
SR on the basis of quantum mechanics (QM), we will have no that for tachyon
either. Then a controversial problem for more than 70 years arose: Do SR and
QM have the common essence? Are they really consistent, compatible or even
identical at the deeper level?\\
\indent In the past, some physicists considered SR being a classical theory
which is different from the QM in their nature and the successful combination
of SR and QM to innovate the relativistic QM (RQM), the quantum field theory
(QFT) and particle physics (PP) is a success of some "complementarity". We
don't think so. We have been pondering on two facts for a long time.\\
\indent First, the derivation of famous formula $\lambda=h/p$ ($p=\hbar k$) by
de 'Broglie in 1924 is by no means a simple combination of SR with the quantum
theory of early stage at his time. What de 'Broglie did is to derive the
"half" of quantum theory $p=\hbar k$ from another "half" of quantum theory
$E=\hbar\omega$ by using the whole theory of SR. His thinking clearly showed
that SR and quantum theory do have the common essence[7]. The second fact
comes from biology: The combination of living beings of different species
cannot reproduce their descendants. It would be interesting to mention an
exotic example. The mating of a horse and a donkey gives birth of a mule, but
the latter can no longer have descendant. Now we know that the genetic factor
in inheritance is DNA. So our problem becomes the following: What are the DNA
of RQM, QFT and PP inherited from SR and QM respectively?\\
\indent In searching for the answer to above question, we have been gradually
accumulating ten arguments for it[7], focusing on one relativistic principle
(as the substitution of original two postulates) --- a basic symmetry in one
inertial system --- based on QM as follows.\\
\indent A particle is always not pure in the sense of its wavefunction (WF)
always containing two contradictory fields, $\varphi(x,t)$ and $\chi(x,t)$.
They evolve with space and time essentially with opposite phase, e.g., in free
motion[8]:
\begin{equation}
\varphi\sim\exp\left[\frac{i}{\hbar}(\vec{p}\cdot\vec{x}-Et)\right],
\end{equation}
\begin{equation}
\chi\sim\exp\left[-\frac{i}{\hbar}(\vec{p}\cdot\vec{x}-Et)\right].
\end{equation}
But in the WF $\psi$ of a concrete particle, $\varphi$ dominates $\chi$, so
the latter has to follow the space-time evolution of the former:
\begin{equation}
\psi\sim\varphi\sim\chi\sim\exp\left[\frac{i}{\hbar}(\vec{p}\cdot\vec{x}-Et)\right],
\hskip 0.3in (|\varphi|>|\chi|).
\end{equation}
By contrast, in the WF $\psi_c$ of its antiparticle, because now $\chi_c$
dominates $\varphi_c$, so:
\begin{equation}
\psi_c\sim\chi_c\sim\varphi_c\sim\exp\left[-\frac{i}{\hbar}(\vec{p}\cdot\vec{x}-Et)\right],
\hskip 0.3in (|\chi_c|>|\varphi_c|).
\end{equation}
Under the newly defined space-time inversion ($\vec{x}\longrightarrow -\vec{x}$,
$t\longrightarrow -t$), the theory (equation) remains invariant but with its
concrete solution transforming from a particle to its antiparticle due to:
\begin{equation}
\left\{
\begin{array}{l}
\varphi(\vec{x},t)\longrightarrow\varphi(-\vec{x},-t)\longrightarrow\chi(\vec{x},t), \\
\chi(\vec{x},t)\longrightarrow\chi(-\vec{x},-t)\longrightarrow\varphi(\vec{x},t).
\end{array}
\right.
\end{equation}
Note that the operators of momentum and energy for a particle:
\begin{equation}
\hat{\vec{p}}=-i\hbar\nabla, \hskip 0.3in \hat{E}=i\hbar\frac{\partial}{\partial t}
\end{equation}
are now transformed into that for its antiparticle:
\begin{equation}
\hat{\vec{p}}_c=i\hbar\nabla, \hskip 0.3in \hat{E}_c=-i\hbar\frac{\partial}{\partial t}.
\end{equation}
So the observable momentum and energy of the particle (15) or its antiparticle
(16) are all the same $\vec{p}$ and $E(>0)$ respectively. Operators (18) and
(19) are exactly the DNA in the RQM, QFT and PP inherited from QM and SR
respectively.\\
\indent Now we are in a position to establish the quantum equation for tachyon
(carrying proper-mass $m_s$) in comparison with Dirac equation (carrying
rest-mass $m_0$)[12]:
$$
Dirac \hskip 0.1in equation \hskip 1.5in Tachyon \hskip 0.1in equation
$$
\begin{eqnarray}
\left\{
\begin{array}{l}
i\hbar\frac{\partial}{\partial t}\varphi=ic\hbar\vec{\sigma}\cdot\nabla\chi+m_0c^2\varphi\\
i\hbar\frac{\partial}{\partial t}\chi=ic\hbar\vec{\sigma}\cdot\nabla\varphi-m_0c^2\chi,
\end{array}
\right.
\ \ \ \ \  &
\left\{
\begin{array}{l}
i\hbar\frac{\partial}{\partial t}\varphi=ic\hbar\vec{\sigma}\cdot\nabla\chi+m_sc^2\chi\\
i\hbar\frac{\partial}{\partial t}\chi=ic\hbar\vec{\sigma}\cdot\nabla\varphi-m_sc^2\varphi,
\end{array}
\right.
\end{eqnarray}
\begin{eqnarray}
\varphi=\frac{1}{\sqrt{2}}(\phi_L+\phi_R), \hskip 0.2in
\chi=\frac{1}{\sqrt{2}}(\phi_L-\phi_R),
\ \ \ \ \  &
\varphi=\frac{1}{\sqrt{2}}(\phi_L+\phi_R), \hskip 0.2in
\chi=\frac{1}{\sqrt{2}}(\phi_L-\phi_R),
\end{eqnarray}
\begin{eqnarray}
\left\{
\begin{array}{l}
i\hbar\frac{\partial}{\partial t}\phi_L=ic\hbar\vec{\sigma}\cdot\nabla\phi_L+m_0c^2\phi_R\\
i\hbar\frac{\partial}{\partial t}\phi_R=-ic\hbar\vec{\sigma}\cdot\nabla\phi_R+m_0c^2\phi_L,
\end{array}
\right.
\ \ \ \ \  &
\left\{
\begin{array}{l}
i\hbar\frac{\partial}{\partial t}\phi_L=ic\hbar\vec{\sigma}\cdot\nabla\phi_L-m_sc^2\phi_R\\
i\hbar\frac{\partial}{\partial t}\phi_R=-ic\hbar\vec{\sigma}\cdot\nabla\phi_R+m_sc^2\phi_L,
\end{array}
\right.
\end{eqnarray}
\begin{eqnarray}
\frac{\partial}{\partial t}\rho+\nabla\cdot\vec{j}=0,
\ \ \ \ \  &
\frac{\partial}{\partial t}\rho+\nabla\cdot\vec{j}=0,
\end{eqnarray}
\begin{eqnarray}
\rho=\varphi^\dag\varphi+\chi^\dag\chi=\phi_L^\dag\phi_L+\phi_R^\dag\phi_R,
\ \ \ \ \  &
\rho=\varphi^\dag\chi+\chi^\dag\varphi=\phi_L^\dag\phi_L-\phi_R^\dag\phi_R,
\end{eqnarray}
\begin{eqnarray}
\begin{array}{lll}
\vec{j} & = & -c(\varphi^\dag\vec{\sigma}\chi+\chi^\dag\vec{\sigma}\varphi) \\
 & = & -c(\phi_L^\dag\vec{\sigma}\phi_L-\phi_R^\dag\vec{\sigma}\phi_R).
\end{array}
\ \ \ \ \  &
\begin{array}{lll}
\vec{j} & = & -c(\varphi^\dag\vec{\sigma}\varphi+\chi^\dag\vec{\sigma}\chi) \\
 & = & -c(\phi_L^\dag\vec{\sigma}\phi_L+\phi_R^\dag\vec{\sigma}\phi_R).
\end{array}
\end{eqnarray}
Here we choose the $\vec{\sigma}$ matrix as the minus of the original one in
Dirac equation such that its counterpart --- the tachyon equation can describe
a left-handed neutrino together with a right-handed antineutrino.\\
\indent Actually, to author's knowledge, the above tachyon equation was found
earlier by Chodos et al.[9], Ciborowski and Rembielinski[10] and Chang[11] by
different approaches but in four-component form. However, we prefer to use the
two-component spinor ($\varphi$, $\chi$ or $\phi_L$, $\phi_R$), showing the
symmetries clearly[12].\\
\indent First, both Dirac equation and tachyon equation respect the basic
symmetry, i.e., the invariance of space-time inversion (17) explicitly. (There
is merely an exchange of $\varphi$ and $\chi$ in the mass term.) A simple
substitution of plane WF to (20) yields the desired kinematic relation (9)
immediately.\\
\indent The tachyon velocity $u$ is identified to the group velocity $u_g$ of
wave:
\begin{equation}
u=u_g=\frac{d\omega}{dk}=\frac{dE}{dp}=\frac{\vec{p}c^2}{E}>c,
\end{equation}
and Eq.(10) follows accordingly.\\
\indent Then next question arises: Although it is important to find both
Dirac equation and tachyon equation obeying the same basic symmetry, but what
is their distinction?

\vskip 48pt
\baselineskip 24pt
\noindent
{\bf 4. What is the Convincing Evidence for Neutrino Being Likely a Tachyon?}

\vskip 24pt
\baselineskip 24pt
\indent The discussion on tachyon is by no means an academic problem as that
before 1960s. Since 1970s, a series of experiments like the beta decay of
tritium show that the neutrino might have nonzero mass $m_\nu$, which is
defined by Eq.(1):
\begin{equation}
E^2=p^2c^2+m_\nu^2c^4.
\end{equation}
Because of great difficulty in the measurement of neutrino's energy $E$,
especially its momentum $p$, the accuracy of experiments has been quite low.
However, in 1996, it was reported that the global averaged value of $m_\nu^2$
for electron-neutrino seemed to be negative[13]:
\begin{equation}
m^2(\nu_e)=-27\pm 20(eV^2).
\end{equation}
Later, the experimental technique was improved to control better the energy
loss of beta particle in the source and nine data in 1991-1995 were excluded
because they were judged as unreliable, resulting in new average in 2000
as[13]:
\begin{equation}
m^2(\nu_e)=-2.5\pm 3.3(eV^2).
\end{equation}
Similar situation occurs for the muon-neutrino[13]:
\begin{equation}
m^2(\nu_\mu)=-0.016\pm 0.023(MeV^2).
\end{equation}
The tau-neutrino was just discovered in FermiLab in 2000, no experimental data
on its $m^2(\nu_\tau)$ is reported yet.\\
\indent To author's knowledge, the majority of physics community don't pay
enough attention to the minus sign in Eqs.(28)-(30), which are regarded
meaningless in statistics (so the present data are still compatible with
$m_\nu^2=0$). Alternatively, some physicists believe that even if neutrino
does have some tiny mass, it must be the rest-mass of a Dirac particle (i.e.,
$m_\nu^2>0$). Only a few papers like [9], [6], [10], [14] and [11] etc. were
considering that the experimental data of negative $m_\nu^2$, though far from
accurate, do strongly hint the neutrino being a tachyon as discussed in
Eqs.(9) and (10) with a real proper mass $m_s$ ($m_\nu^2=-m_s^2$).\\
\indent We hope that the experimental accuracy could be raised in the near
future so that whether the neutrino is a tachyon or not will be verified
unambiguously. However, in our point of view, the convincing evidence for the
neutrino being a tachyon is already lying in the fact of parity
violation[12, 15, 16].\\
\indent Since the historical discovery of parity violation by Lee-Yang[17],
and Wu et al.[18], all decay processes in weak interaction where neutrinos
participate in are explained by the two-component neutrino theory[19], which
implies that there are only left-handed neutrino and right-handed antineutrino
in nature whereas no right-handed neutrino or left-handed antineutrino exists.
What renders this kind of permanent longitudinal polarization of neutrino
possible is its mass being regarded as zero at that time. Would the neutrino
have a nonzero rest-mass, no matter how tiny it is, the neutrino, say a
$\nu_L$, would be a Dirac particle with velocity $u<c$ in a system $S$. Then
an another observer in $S'$ system moving with respect to $S$ with a velocity
$v>u$ would see the $\nu_L$ as a $\nu_R$! There would be no permanent
longitudinal polarization at all. We would still need a four-component
neutrino theory with parity conservation. This obviously contradicts the
experimental facts of parity violation. Some physicists think that even the
neutrino has some rest-mass, the experimental facts of parity violation could
still be accounted for by the V-A theory in the standard model which is
endowed with the property of parity violation. We don't think so. Would
particles in weak interaction process be all Dirac particles, there would be
no parity violation phenomenon at all. In particular, one even cannot
discriminate a static fermion being left-handed or right-handed polarized[20].
This is because the solution to the Dirac equation describing a plane wave
along $z$ axis reads:
\begin{equation}
\psi\sim\varphi\sim\chi\sim\phi_L\sim\phi_R\sim\exp\left[\frac{i}{\hbar}(pz-Et)\right],
\end{equation}
which will give $\phi_L=\phi_R$ if $p=0$.\\
\indent On the other hand, in the two-component neutrino theory, the Dirac
equation with $m_0=0$ reduces into two uncoupled Weyl equations. Only one of
them:
\begin{equation}
i\hbar\frac{\partial}{\partial t}\phi_L=ic\hbar\vec{\sigma}\cdot\nabla\phi_L
\end{equation}
is adopted whereas the other one:
\begin{equation}
i\hbar\frac{\partial}{\partial t}\phi_R=-ic\hbar\vec{\sigma}\cdot\nabla\phi_R
\end{equation}
is discarded. Eq.(32) precisely describes the massless $\nu_L$ and
$\bar{\nu}_R$.\\
\indent Note that the WF of a $\bar{\nu}_R$ moving along $z$ axis reads:
\begin{equation}
\phi_L\sim\left(
\begin{array}{c}
0\\
1
\end{array}
\right)\exp\left[-\frac{i}{\hbar}(pz-Et)\right], \hskip 0.3in
(E>0, \hskip 0.1in p>0).
\end{equation}
It is a "negative-energy" neutrino, but actually a positive-energy
antineutrino with momentum $p$ and energy $E>0$ if we use the correct
operators for antiparticle, Eq.(19). Meanwhile, the spin angular-momentum
operator also reads:
\begin{equation}
\hat{\vec{\sigma}}_c=-\vec{\sigma},
\end{equation}
as can be proved by the Heisenberg motion equation. So Eq.(34) exactly
describes a right-handed antineutrino without any resort to so-called
"hole-theory". It was emphasized by Konopinski and Mahmaud[21], Schwinger[22]
even earlier by Stueckelberg and Feynman in different form and to different
extent, (see below).\\
\indent Now we turn to the tachyon equation (20)-(22) with nonzero proper mass
$m_s$. In contrast to Dirac equation, it is not invariant under the following
space-inversion transformation ($\vec{x}\longrightarrow -\vec{x}$,
$t\longrightarrow t$):
\begin{equation}
\left\{
\begin{array}{l}
\phi_L(\vec{x}, t)\longrightarrow\phi_L(-\vec{x}, t)\longrightarrow\phi_R(\vec{x}, t)\\
\phi_R(\vec{x}, t)\longrightarrow\phi_R(-\vec{x}, t)\longrightarrow\phi_L(\vec{x}, t),
\end{array}
\right.
\end{equation}
because of the opposite signs in its mass term. This is a violation to parity
symmetry and must be reflected in its solutions. Indeed, consider a plane wave
like
\begin{equation}
\psi\sim\varphi\sim\chi\sim\phi_L\sim\phi_R\sim\exp\left[\frac{i}{\hbar}(pz-Et)\right],
\end{equation}
which gives:
\begin{equation}
\chi=\frac{cp-m_sc^2}{E}\varphi,
\end{equation}
\begin{equation}
\phi_R=\frac{m_sc^2}{cp+E}\phi_L.
\end{equation}
We see that for $E>0$, a paticle always have $|\varphi|>|\chi|$ and
$|\phi_L|>|\phi_R|$, i.e., it is a left-handed neutrino $\nu_L$. But for $E<0$,
an antiparticle always have $|\chi_c|>|\varphi_c|$ and $|\phi_R|>|\phi_L|$,
i.e., it is a right-handed antineutrino $\bar{\nu}_R$ with moment $p_c=-p$ and
energy $E_c=-E>0$. Moreover, the solution of $\nu_R$ or $\bar{\nu}_L$ is
definitely excluded. The above feature can also be seen from the expression
for the "charge density", Eq.(24), being not positive definite: While the
normalization for a particle $\nu_L$, $\int\rho d\vec{x}=1$ implies the
helicity $H=<\vec{\sigma}\cdot\vec{p}>/|\vec{p}|=-1$, it will equal to -1 for
an antiparticle $\bar{\nu}_R$ and implies $H=1$.\\
\indent In summary, only a neutrino with intrinsic left-right asymmetry can
account for the parity violation in weak interactions. And only a neutrino
with $m_\nu=0$ or $m_\nu^2<0$ can be such an initiator of parity violation.
The reason why we believe $m_\nu^2<0$ rather than $m_\nu=0$ is not only due to
the present experimental data tending to $m_\nu^2<0$, but largely depending on
a theoretical confidence that "a particle is always not pure and there is no
exception to this rule". So a neutrino should be a superposition of two fields
$\phi_L$ and $\phi_R$, rather than one field. Just look at all known particles.
Even the massless photon, despite its special character, is no exception. A
left-handed and right-handed photon can be viewed as a particle and its
antiparticle respectively. But a linearly polarized light-beam is a
superposition of them.

\vskip 48pt
\baselineskip 24pt
\noindent
{\bf 5. The Full Solution to Superluminal Paradox}

\vskip 24pt
\baselineskip 24pt
\indent Now it is easy to solve the superluminal paradox, Eqs.(5), (7), (11)
and (12). Evidently, since the observer in $S'$ system will see the particle
(with velocity $u>c^2/v$ in $S$ system) as a negative-energy ($E'<0$) particle,
he should look it as an antiparticle with positive-energy $E_c=-E'>0$.
Meanwhile, its momentum should be $P_c=-p'<0$ by the formula (19), in
conformity with $u'<0$ in Eq.(7). The mysterious time-reversal Eq.(5) is no
more than a false appearance of sign change in the phase of WF, which, of
course, cannot be reflected suitably in the "classical" $x-t$ diagram. So all
puzzles disappear now. There is no paradox at all.\\
\indent However, we have to check carefully if the above "reinterpretation
rule" is really allowed by the equation? Indeed, it does work for a neutrino
described by Eq.(22) in $S'$ system, which is invariant under the "pure
time-inversion" ($\vec{x}'\longrightarrow\vec{x}'$, $t'\longrightarrow -t'$)
as far as :
\begin{equation}
\left\{
\begin{array}{l}
\phi_L(\vec{x}', t')\longrightarrow\phi_L(\vec{x}', -t')\longrightarrow\phi_R(\vec{x}', t')\\
\phi_R(\vec{x}', t')\longrightarrow\phi_R(\vec{x}', -t')\longrightarrow\phi_L(\vec{x}', t'),
\end{array}
\right.
\end{equation}
with a concrete solution ($p'>0$, $E'<0$):
\begin{equation}
\phi_L\sim\phi_R\sim\exp\left[\frac{i}{\hbar}(p'x'-E't)\right],
\end{equation}
being transformed into:
\begin{equation}
\phi_R\sim\phi_L\sim\exp\left[-\frac{i}{\hbar}(p_cx'-E_ct)\right],
\end{equation}
which is just an right-handed antineutrino with $p_c=-p<0$, $E_c=-E>0$.
Q.E.D.\\
\indent Note that while neutrino equation (22) allows such kind of a "pure
time-inversion", the Dirac equation cannot. On the other hand, both (all)
equations should be invariant under a full space-time inversion
($x\longrightarrow -x$, $t\longrightarrow -t$) while a concrete particle
transforming into its antiparticle with the same momentum and (positive)
energy accordingly. This is the rigorous statement of stuekelberg-Feynman
rule.\\
\indent Now the momentum of antineutrino in Eq.(41) is negative: $p_c<0$,
opposite to the direction of original $p(>0)$. So the implication is amazing:
when the $S'$ observer chases the neutrino (with velocity $u$ in $S$ system)
and increases his velocity $v$ exceeding a critical value $c^2/u$, the $\nu_L$
will suddenly transform into a $\bar{\nu}_R$ moving toward him[5, 6].

\vskip 48pt
\baselineskip 48pt
\noindent
{\bf 6. Why a Dirac Equation with imaginary mass doesn't work?}

\vskip 24pt
\baselineskip 24pt
\indent  At first sight, the difference between Eqs.(1)-(3) and Eqs.(9)-(10)
amounts simply to an analytical continuation of particle mass:
$m_0\longrightarrow im_s$ (with both $m_0$ and $m_s$ being real). Such kind of
unsuccessful attempt could be found in the literature, e.g., [23].\\
\indent Look at Dirac equation versus Eq.(20). In terms of four-component form,
they are:
\begin{equation}
i\hbar\frac{\partial}{\partial t}\psi=ic\hbar\vec{\alpha}\cdot\nabla\psi
+\beta m_0c^2\psi,
\end{equation}
\begin{equation}
i\hbar\frac{\partial}{\partial t}\psi=ic\hbar\vec{\alpha}\cdot\nabla\psi
+\beta_sm_sc^2\psi,
\end{equation}
\begin{equation}
\vec{\alpha}=\left(
\begin{array}{cc}
0 & \vec{\sigma}\\
\vec{\sigma} & 0
\end{array}
\right), \hskip 0.3in
\beta=\left(
\begin{array}{cc}
I & 0\\
0 & -I
\end{array}
\right), \hskip 0.3in
\beta_s=\left(
\begin{array}{cc}
0 & I\\
-I & 0
\end{array}
\right).
\end{equation}
While $\beta$ is hermitian, $\beta_s$ is antihermitian but it works. On the
other hand, a direct analytical continuation $m_0\longrightarrow im_s$ leads
to Eq.(43) with
\begin{equation}
\beta_s\longrightarrow\beta'_s=\left(
\begin{array}{cc}
iI & 0\\
0 & -iI
\end{array}
\right),
\end{equation}
which is also antihermition. Why it is wrong?\\
\indent As pointed out in [15], usually in QM, we endow an imaginary part of
mass $m$ to WF for descrbing an unstable particle:
\begin{equation}
m\longrightarrow m-i\Gamma/2, \hskip 0.3in \Gamma=\hbar/\tau, \hskip 0.3in
|\psi|^2\sim e^{-t/\tau},
\end{equation}
which implies the unitarity of WF being destroyed. So it is no surprise to see
that the violation of hermitian property in Eqs.(43) with (45) allows a
decoupled solution at $p\longrightarrow 0$ like:
\begin{equation}
\varphi'\sim e^{-i(im_st)}\sim e^{m_st}
\end{equation}
\begin{equation}
\chi'\sim e^{-i(-im_st)}\sim e^{-m_st}
\end{equation}
which implies a violation of unitarity too.\\
\indent The reason why Eq.(43) is correct can be seen from a simple but new
derivation of Eq.(22) (in Weyl representation) from Dirac equation via a
nonhermitian unitary transformation[15]:
\begin{equation}
\psi'=\left(
\begin{array}{c}
\phi'_L\\
\phi'_R
\end{array}
\right)\longrightarrow U_s\psi'=\left(
\begin{array}{cc}
iI & 0\\
0 & I
\end{array}
\right)\psi'=\psi=\left(
\begin{array}{c}
\phi_L\\
\phi_R
\end{array}
\right).
\end{equation}
Thus we see that the extra $i$ coming from the analytical continuation of mass
is now absorbed into the definition of $\phi_L=i\phi'_L$ (while $\phi_R=\phi'_R$).
The latter implies an extra phase-difference between $\phi_L$ and $\phi_R$ (in
comparison with $\phi'_L$ and $\phi'_R$) which ensures two solutions (for a
same momentum) $\nu_L$ and $\bar{\nu}_R$ being stabilized whereas other two
$\nu_R$ and $\bar{\nu}_L$ strictly forbidden. Moreover, the solution like (47)
is definitely excluded. In short, the violation of hermitian property for an
equation (usually leading to the violation of unitarity) is now displayed in a
stable but strange realization of parity violation. We see once again the
subtlety of phase in QM.

\vskip 48pt
\baselineskip 24pt
\noindent
{\bf 7. Does So-called Pseudotachyon Exist?}

\vskip 24pt
\baselineskip 24pt
\indent In Ref.[2], a so-called "Pseudotachyon" was proposed that it behaves
like tachyons in the momentum space but like subluminal particle in the
ordinary space. How can one believe in such strange property which would lead
to decoupling of particle from wave? Let's study this puzzle carefully.\\
\indent First, the kinematic relation (9) in momentum space leads
straightforwardly to the group velocity of a tachyon as shown by Eq.(26).\\
\indent Second, in the Heisenberg picture, one can use the motion equation for
the position operator $x$ to find the velocity operator for Dirac equation
being:
\begin{equation}
\dot{x}_i=\frac{1}{i\hbar}[x_i,\ H]=c\alpha_i, \hskip 0.3in (i=1,2,3)
\end{equation}
where $H=c\vec{\alpha}\cdot\vec{p}+\beta m_0c^2$. The motion equation for
$\alpha_i$ yields further the solution of $x$ being:
\begin{equation}
x=x_0+c^2pH^{-1}t+\frac{1}{2}i\hbar(\dot{x})_0H^{-1}e^{-2iHt/\hbar},
\end{equation}
where $c^2pH^{-1}$ is just the classical velocity shown in (26). The last term
was named the "zitterbewegung", which is complex and thus unobservable. Hence
the position operator $x$ ceases to be an observable in RQM[24]. But anyway,
the velocity or position of a Dirac particle does make sense in the average
meaning. The above evaluation remains valid for a tachyon because it does not
alter after $\beta$ being replaced by $\beta_s$ in (44).\\
\indent Third, we turn to the Schr\"{o}dinger picture and calculate the
average velocity of a Dirac particle:
\begin{equation}
\bar{v}=c\int\psi^{\dag}\alpha_3\psi d\vec{x}=c^2p/E<c
\end{equation}
where a planewave along $z$ axis is used and the WF is normalized in a volume
$V$ such that
\begin{equation}
\int\psi^{\dag}\psi d\vec{x}=1
\end{equation}
In [2], the same formula was used to calculate the average velocity of a
tachyon, yielding the strange but wrong result:
\begin{equation}
\bar{v}=E/p<c
\end{equation}
What's wrong? While Eqs.(52) and (51) are correct for Dirac equation, we
should return back to its starting point as shown in Eqs.(24) and (25). The
tachyon equaton has its own $\rho$ and $\vec{j}$ which must be derived from
the continuity equation and are quite different from that of Dirac equation.
Indeed, using $\int\rho d\vec{x}=1$ to find the correct normalization constant
for a plane-wave of tachyon along $z$ axis and evaluating $j_3$, we arrive at:
\begin{equation}
\bar{v}=c^2p/E=u>c
\end{equation}
as expected. In short, the correct expressions for tachyon should be Eqs.(52)
and (51) with an extra matrix
$\gamma_5=\left(
\begin{array}{cc}
0 & I\\
I & 0
\end{array}
\right)$ inserted.\\
\indent As three different approaches --- the group velocity, the Heisenberg
motion equation and the average velocity in Schr\"{o}dinger picture --- are
all focusing on one outcome that a tachyon does travel in a superluminal speed
$u=c^2p/E>c$, we have more confidence than before.

\vskip 48pt
\baselineskip 24pt
\noindent
{\bf 8. Summary and Discussion}

\vskip 24pt
\baselineskip 24pt
\indent (a) Numerous experimental tests have been supporting the validity of
SR, which stands even more firm than ever before. However, it is possible to
construct a superluminal theory compatible with SR. Indeed, both subluminal
and superluminal quantum theories are respecting the same basic symmetry,
i.e., the invariance under the (newly defined) space-time inversion, Eq.(17).\\
\indent (b) According to our present understanding, no boson but fermion could
be superluminal as long as the parity symmetry, i.e., the invariance under the
space inversion, Eq.(36), is violated to maximum. The present experimental
data, especially that of parity violation, do strongly hint that the neutrino
might be such a superluminal particle, i.e., a tachyon with permanent
longitudinal polarization.\\
\indent (c) Interesting enough, the Lorentz transformation (LT) and the
addition law for velocity are valid for both subluminal and superluminal
phenomena as long as the relative velocity $v$ of observers in two inertial
systems is constrained: $|v|<c$. All of our discussion can be meaningful only
if it is based on SR, LT and the invariance of light-speed $c$.\\
\indent (d) The superluminal paradox is over. All puzzles stemming from the
classical concept disappear in the reasonable quantum theory. Indeed, the
solution to superluminal paradox poses a very severe and interesting test on
the validity of Eq.(17) which in turn is based on the new concept about the
symmetry between a particle and its antiparticle (including Eq.(19)). Now we
have a sound basis for the "Stuekelberg-Feynman-Sudarshan-Recami
reinterpretation rule".\\
\indent (e) Five topics on the research of neutrino were discussed in Ref.[16].
The puzzle of neutrino oscillation seems more interesting. Either Eq.(1) or
Eq.(9) yields a simple relation between the group velocity $u_g=\frac{d\omega}{dk}$
and the phase velocity $u_p=\frac{\omega}{k}$ as:
\begin{equation}
u_gu_p=c^2.
\end{equation}
Hence, if neutrinos ($\nu_e$, $\nu_{\mu}$ and $\nu_{\tau}$) are really
tachyons with $u_g>c$ but $u_p<c$, they would be difficult to form coherent
superposing state during their motion. So the probability of flavor mixing,
i.e., the oscillation among them would be strongly suppressed. We guess that
it might be one of the reasons why the present fitting value for the rest-mass
of neutrino from the experimental data of oscillation is so tiny in comparison
with the data shown in Eqs.(28)-(30).\\
\indent (f) In the development from nonrelativistic QM to RQM, the position
and velocity operators cease to be hermitian, i.e., they do not correspond to
direct observables but only make sense in the average meaning. Besides, as is
well known, the Klein-Gordon equation has its Hamiltonian being
nonhermitian[25] and its "probability density" ($\rho$) being not positive
definite[8,24]. Now we see similar situation occured in the tachyon equation
too. All above features in RQM are intimately linked with the antiparticle's
degree of freedom --- a particle is always composed of two contradictory
fields, not one field.

\vskip 48pt
\baselineskip 24pt
\noindent
{\bf Acknowledgments}

\vskip 24pt
\baselineskip 24pt
\indent The author is grateful to T. Chang and E. Recami for bringing
Refs.[2] and [1,6] to his attention respectively.

\vskip 48pt
\baselineskip 24pt
\noindent
{\bf References}

\vskip 24pt
\baselineskip 24pt
\noindent
[1] E. Recami, {\it Foundation of Phys.} {\bf 31}, 1119 (2001).\\
\noindent
[2] G. Salesi, {\it Int. J. Mod. Phys.} A {\bf 12}, 5103 (1997).\\
\noindent
[3] O.M. Bilaniuk, V.K. Deshpande and E.C.G. Sudarshan, {\it Am. J. Phys.}
{\bf 30}, 718 (1962).\\
\noindent
[4] O.M. Bilaniuk et al. {\it Phys. Today} {\bf 22}, May 43; Dec. 47 (1969).\\
\noindent
[5] G-j Ni, {\it Superluminal paradox and neutrino,} preprint, hep-ph/0203060.\\
\noindent
[6] E. Giannetto, G.D. Maccarrone, R. Mignani, E. Racami, {\it Phys. Lett.}
{\bf B178}, 115 (1986).\\
\noindent
[7] G-j Ni, {\it Ten arguments for the essence of special relativity},
Proceedings of the 23rd Workshop on High-energy Physics and Field
Theory, (Protvino, Russia, June, 2000) p.275-292.\\
\noindent
[8] G-j Ni, H. Guan, W-m Zhou, J. Yan, {\it Chin. Phys. Lett.} {\bf 17}, 393
(2000).\\
\noindent
[9] A. Chodos, A.I. Hauser and V.A. Kostelecky, {\it Phys. Lett.} {\bf B150},
431 (1985).\\
\noindent
[10] J. Ciborowski and J. Rembielinski, {\it Europ. Phys. J.} {\bf C8}, 157
(1999).\\
\noindent
[11] T. Chang and G-j Ni, {\it An explanation on negative mass-square of
neutrinos}, accepted by Fizika B, hep-ph/0009291.\\
\noindent
[12] G-j Ni and T. Chang, {\it Is neutrino a superluminal particle?} preprint,
hep-ph/0103051.\\
\noindent
[13] Review of Particle Physics, {\it Phys. Rev.} {\bf D54}, 280 (1996);
{\it Europ. Phys. J.} {\bf C15}, 350 (2000).\\
\noindent
[14] T. Chang, {\it Beyond relativity}, preprint (2000).\\
\noindent
[15] G-j Ni, {\it J. Shaanxi Normal Univ.} (Natural Sci. Ed.) {\bf 29}, (3) 1
(2001).\\
\noindent
[16] G-j Ni, {\it WULI (physics)}, April (4) 255 (2002).\\
\noindent
[17] T.D. Lee and C.N. Yang, {\it Phys.Rev.} {\bf 104}, 254 (1956).\\
\noindent
[18] C.S. Wu et al. {\it Phys.Rev.} {\bf 105}, 1413 (1957).\\
\noindent
[19] T.D. Lee and C.N. Yang, {\it Phys.Rev.} {\bf 105}, 1671 (1957).\\
\noindent
[20] L.H. Ryder, {\it Quantum Field Theory} (Cambridge, Cambridge Univ.
Press, 1996).\\
\noindent
[21] E.J. Konopinski and H.M. Mahmaud, {\it Phys.Rev.} {\bf 92}, 1045 (1953).\\
\noindent
[22] J. Schwinger, {\it Proc. Nat. Acad. Sc. US}, {\bf 44}, 223 (1958).\\
\noindent
[23] J. Bandukwala and D. Shay, {\it Phys.Rev.} {\bf D9}, 889 (1974).\\
\noindent
[24] G-j Ni and S-q Chen, {\it Advanced Quantum Mechanics} (Press of Fudan
University, 2000); The English version will be published by Rinton Press in
2002.\\
\noindent
[25] H. Feshbach and F. Villars, {\it Rev. Mod. Phys.} {\bf 30}, 24 (1958).
\end{document}